# Polarizabilities of an Annular Cut in the Wall of an Arbitrary Thickness

Sergey S. Kurennoy

*Abstract*— The electric and magnetic polarizabilities of an aperture are its important characteristics in the theory of aperture coupling and diffraction of EM waves. The beam coupling impedances due to a small discontinuity on the chamber wall of an accelerator can also be expressed in terms of the polarizabilities of the discontinuity. The polarizabilities are geometrical factors which can be found by solving a static (electric or magnetic) problem. However, they are known in an explicit analytical form only for a few simple-shaped discontinuities, such as an elliptic hole in a thin wall. In the present paper the polarizabilities of a ring-shaped cut in the wall of an arbitrary thickness are studied using a combination of analytical, variational and numerical methods. The results are applied to estimate the coupling impedances of button-type beam position monitors.

## I. Introduction

IN the theory of diffraction and penetration of EM waves through apertures in conducting walls many important quantities can be related to the aperture polarizabilities [1], [2], [3]. The coupling impedances of a small discontinuity on the wall of the vacuum chamber of an accelerator have also been calculated in terms of the polarizabilities of the discontinuity [4]. The basic idea of the approach used is due to the Bethe theory of diffraction by small holes [1], which shows that the fields scattered by a hole can be approximated by those produced by effective dipoles which are induced on the hole by an incident (beam or incident-wave) field. The magnitudes of the effective electric $P$ and magnetic $M$ dipoles are expressed through the incident fields $E_\nu^h, H_\tau^h$ at the hole location without hole [1], [2]

$$P_\nu = -\chi \varepsilon_0 E_\nu^h/2; \quad M_\tau = \psi H_\tau^h/2 , \qquad (1)$$

where $\chi$ is the electric polarizability and $\psi$ is the magnetic susceptibility of the hole, $\hat{\nu}$ is the normal vector to the hole plane, and $\hat{\tau}$ is the tangential one. In general, $\psi$ is a two-dimensional (2-D) symmetric tensor, but we restrict ourselves here by only axisymmetric holes.

When the wavelength of an incident field is large compared to a typical size of the aperture, the aperture is considered to be small and its polarizabilities can be found by solving an electrostatic or magnetostatic problem [2]. The solutions are known in an explicit analytical form for a few simple cases, see review [3]. For a circular hole of radius $b$ in a zero-thickness wall $\psi = 8b^3/3$ and $\chi = 4b^3/3$ [1]. Analytical results for elliptic holes in a thin wall are also available [2]. In the case of a thick wall the polarizabilities have been studied using a variational technique for circular holes [5] and for elliptic holes [6]. Some approximate formulas for slots are compiled in [7].

In the present paper, we consider an annular cut in the perfectly conducting planar wall of an arbitrary thickness. In other words, the geometry under investigation is a circular hole in the wall with a concentric disk placed in it. This aperture can serve as a model of a coax attached to the waveguide when the wall thickness is large enough. In the case of a thin or finite-thickness wall it is an approximation of an electrode of the button-type beam position monitors (BPMs). For this geometry, an integral equation for electric and magnetic potential is derived in Section II. The magnetic problem is studied in Section III. The integral equation is first solved analytically for a narrow cut in a thin wall, and then studied by variational methods in other cases. For the electric problem, analytical estimates are given and then a numerical approach is used in Section IV. In Section V the results are applied to estimate the beam coupling impedances of BPMs.

## II. General Analysis

### A. Problem symmetry

When the wavelength is large compared to the hole size, the polarizabilities can be obtained by solving the following problem: to find the field distribution produced by the aperture (hole) in a metal planar wall when it is illuminated from one side by a homogeneous static (normal electric or tangential magnetic) field.

Suppose the midplane of a conducting wall of thickness $t$ is at $z = 0$ so that the wall surfaces are in planes $z = \pm t/2$. The center of a hole in the wall coincides with the origin of the plane coordinates $(u, v)$. Let the hole be illuminated by a homogeneous electric field $E_0$ from $z > 0$ side, directed along the normal $-\hat{z}$ to the wall. In the magnetic case, a homogeneous tangential magnetic field $H_0$ is assumed to be directed along $\hat{u}$. It is convenient, following [8], [5], to split the problem into its symmetric and antisymmetric parts, with respect to the corresponding potential. For this purpose, we decompose the far field as $E_0/2 + E_0/2 = E_0$ for $z > 0$, and as $E_0/2 - E_0/2 = 0$ for $z < 0$, and consider two separate problems: (i) the wall with the aperture is immersed into homogeneous field $E_0/2$ — the antisymmetric problem for the electrostatic potential with respect to reflection $z \to -z$; and (ii) the aperture in the wall is illuminated by the field directed to the wall from both sides, $E_0/2$ for $z > 0$ and $-E_0/2$ for $z < 0$, in which case the potential is symmetric.

In a general case, solving separately symmetric and antisymmetric parts of the electric problem yields two polar-

The author is with the Physics Department, University of Maryland, College Park, MD 20742, USA.

izabilities $\chi_s$ and $\chi_a$. They, in turn, give us the inside electric polarizability $\chi_{in} = \chi_s + \chi_a$, which defines the effective electric dipole for the illuminated side of the wall, $z > t/2$, and the outside one, $\chi_{out} = \chi_s - \chi_a$, for the shadow side of the wall, $z < -t/2$. Likewise, the magnetic polarizabilities are $\psi_{in} = \psi_s + \psi_a$ and $\psi_{out} = \psi_s - \psi_a$. For a zero-thickness plane, obviously, the antisymmetric problem is trivial (the field is $E_0/2$ or $H_0/2$ everywhere), so that $\chi_a$ and $\psi_a$ are both zero.

### B. Integral Equations

Let us start from the magnetic problem for a zero-thickness wall, $t = 0$. As mentioned before, the antisymmetric potential is zero everywhere in this case. The symmetric problem can be reduced to the integral equation [8] for the function $G(\vec{r}) = 2H_z(\vec{r}, 0)/H_0$

$$\int_h d\vec{r}' G(\vec{r}') K(\vec{r}, \vec{r}') = u , \qquad (2)$$

where $\vec{r} = (u, v)$, the integration runs over the aperture, and the kernel is symmetric

$$K(\vec{r}, \vec{r}') = \frac{1}{4\pi^2} \int \frac{d\vec{\sigma}}{\sigma} e^{i\vec{\sigma}(\vec{r}-\vec{r}')} = \frac{1}{2\pi |\vec{r} - \vec{r}'|} . \qquad (3)$$

If Eq. (2) is solved, the magnetic susceptibility is [8]

$$\psi_u = \int_h d\vec{r}' u G(\vec{r}') . \qquad (4)$$

For an axisymmetric aperture, one can simplify Eq. (2) using $u = r \cos\varphi$, substituting $G(\vec{r}) = g(r) \cos\varphi$, and intergating over the polar angle $\varphi'$. It yields

$$\int_{[h]} dr' r' g(r') K_m(r, r') = r , \qquad (5)$$

with the following kernel

$$\begin{aligned} K_m(x, y) &= \int_0^\infty d\sigma J_1(\sigma x) J_1(\sigma y) \qquad (6) \\ &= \theta(y - x) \frac{x}{2y^2} {}_2F_1\left(\frac{3}{2}, \frac{1}{2}; 2; \frac{x^2}{y^2}\right) + \{x \leftrightarrow y\} \\ &= \frac{xy}{2(x+y)^3} {}_2F_1\left[\frac{3}{2}, \frac{3}{2}; 3; \frac{4xy}{(x+y)^2}\right] , \end{aligned}$$

where $J_n(x)$ is the n-th order Bessel function of the first kind, and ${}_2F_1$ is the Gauss hypergeometric function. This kernel has a ln-singularity at $x = y$

$$K_m(x, y) \simeq \frac{8xy}{\pi(x+y)^3} \left(\ln\frac{x+y}{|x-y|} + 2\ln 2 - 2\right) + O(|x-y| \ln|x-y|) . \qquad (7)$$

The magnetic susceptibility in this case is

$$\psi = \pi \int_{[h]} dr r^2 g(r) . \qquad (8)$$

In Eqs. (5) and (8) symbol $[h]$ denotes the interval of the radius-vector variation: $[h] = [0, b]$ for a circular hole of radius $b$, and $[h] = [a, b]$ for an annular cut with inner radius $a$ and outer radius $b$.

For the case of finite thickness $t > 0$, one should consider both the symmetric and antisymmetric problems. In this case an integral equation is derived for the function $G(\vec{r}) = 2H_z(\vec{r}, t/2)/H_0$. For an axisymmetric aperture, the integral equation of the symmetric problem is

$$\int_{[h]} dr' r' g(r') [K_m(r, r') + K_{mt}^s(r, r')] = r , \qquad (9)$$

where the thickness-dependent addition $K_{mt}^s$ to the kernel is related to a field expansion inside the aperture, $|z| < t/2$. For the annular gap with radii $a$ and $b$, it has the form

$$K_{mt}^s(r, r') = \sum_{n=1}^\infty F_n(r) F_n(r') \frac{\tanh(\lambda_n t/2)}{\lambda_n} , \qquad (10)$$

where $\lambda_n$ are subsequent positive roots of the equation

$$J_1'(\lambda_n a) Y_1'(\lambda_n b) - Y_1'(\lambda_n a) J_1'(\lambda_n b) = 0 , \qquad (11)$$

$Y_1(x)$ is the Bessel function of the second kind, and the expansion functions $F_n$ are

$$F_n(r) = C_n \left[ J_1(\lambda_n r) - \frac{J_1'(\lambda_n a)}{Y_1'(\lambda_n a)} Y_1(\lambda_n r) \right] . \qquad (12)$$

These functions are normalized to satisfy the condition $\int_a^b r F_n^2(r) = 1$, which defines

$$\begin{aligned} C_n &= \frac{\pi \lambda_n}{\sqrt{2}} \left\{ [Y_1'(\lambda_n b)]^{-2} \left[1 - (\lambda_n b)^{-2}\right] - \qquad (13) \right. \\ &\qquad \left. [Y_1'(\lambda_n a)]^{-2} \left[1 - (\lambda_n a)^{-2}\right] \right\}^{-1/2} . \end{aligned}$$

Likewise, for the antisymmetric problem the thickness-dependent part $K_{mt}^{as}$ of the kernel replaces $K_{mt}^s$ in the integral equation (9), and it is given by Eq. (10) with replacement $\tanh \to \coth$. We do not provide a detailed derivation of the integral equations above since it is quite analogous to that in Ref. [5] for a circular hole. The only difference is in the form of functions $F_n(r)$ and $\lambda_n$ for the thickness-dependent part.

In a similar way, a solution $f(r)$ of the electrostatic problem for a thin wall satisfies the integral equation

$$\int_{[h]} dr' r' f(r') K_e(r, r') = 1 , \qquad (14)$$

with a more singular $[O((x-y)^{-2})]$ kernel

$$K_e(x, y) = \int_0^\infty d\sigma \sigma^2 J_0(\sigma x) J_0(\sigma y) . \qquad (15)$$

The thickness-dependent parts of the kernel have the form similar to Eq. (10). The electric polarizability of the axisymmetric hole is

$$\chi = 2\pi \int_{[h]} dr r f(r) . \qquad (16)$$

A solution $g(r)$ of the integral equation (5) or (9) must have the correct singular behavior near the metal edge. For a zero-thickness wall, the singularity is $g(r) \propto \Delta^{-1/2}$ when distance from the edge $\Delta = b - r \to 0$ or $\Delta = r - a \to 0$. For the electric problem (14), the function $f(r)$, which is proportional to the electric potential, must behave as $\sqrt{\Delta}$ near the edge to provide for the correct singularity $\Delta^{-1/2}$ of the electric field. In the case of a circular hole of radius $b$ the exact solutions of Eqs. (5) and (14) are known [1]. They are $g(r) = 4r/(\pi\sqrt{b^2 - r^2})$ and $f(r) = 2\sqrt{b^2 - r^2}/\pi$, substituting of which in (8) and (16) gives the polarizabilities of a circular hole cited in Introduction. For a thick wall, the corresponding near-edge behavior is $g(r) \propto \Delta^{-1/3}$ and $f(r) \propto \Delta^{2/3}$, assuming 90° edge.

### III. MAGNETIC PROBLEM

#### A. Narrow Cut in Thin Wall: Analytical Solution

Suppose the width $w = b - a$ of the gap is small, $w \ll b$. Introducing dimensionless variables $x = r'/b$ and $y = r/b$, we are looking for a solution of Eq. (5) in the form $g(x) = C(x)/\sqrt{(1 - x)(x - \rho)}$, where $\rho = a/b$, and $C(x)$ is a regular function in the interval $[\rho, 1]$. For a narrow gap $\delta \equiv 1 - \rho \ll 1$, and one can expand $C(x)$ as $C(x) = C + O(\delta)$. Substituting this into Eq. (5) and keeping only the singular part (7) of the kernel (the rest would give corrections $O(\delta)$ to the RHS) leads to the equation

$$1 = \frac{C}{\pi} \int_\rho^1 \frac{dx \, [\ln(8/|x - y|) - 2]}{\sqrt{(1 - x)(x - \rho)}} \,, \qquad (17)$$

where we neglected terms $O(\delta \ln \delta)$ in the RHS. Replacing variables $x = 1 - u\delta$, $y = 1 - v\delta$, and using the identity

$$\int_0^1 \frac{du \, \ln|u - v|}{\sqrt{u(1 - u)}} = -2\pi \ln 2 \,,$$

we get from (17)

$$C = [\ln(32/\delta) - 2]^{-1} \,. \qquad (18)$$

Then from Eq. (8) the magnetic polarizability of a narrow ($w = b - a \ll b$) annular cut in a thin plate is

$$\psi = \frac{\pi^2 b^2 a}{\ln(32b/w) - 2} \,. \qquad (19)$$

It is interesting to compare Eq. (19) with the estimate [7] obtained by approximating the annular cut with an octagon and using the magnetic susceptibilities for narrow slots:

$$\psi_o = \frac{4}{3} \left(\frac{\pi}{4}\right)^4 \frac{b^3}{\ln(2\pi b/w) - 7/3} \,. \qquad (20)$$

While the behavior is similar, this estimate is a few times smaller than (19), see Fig. 1. Moreover, even a more extreme model — two long slots of length $2b$ and width $w$ oriented parallel to the magnetic field — give the polarizability

$$\psi_m = \frac{4}{3} \frac{\pi b^3}{\ln(16b/w) - 7/3} \,, \qquad (21)$$

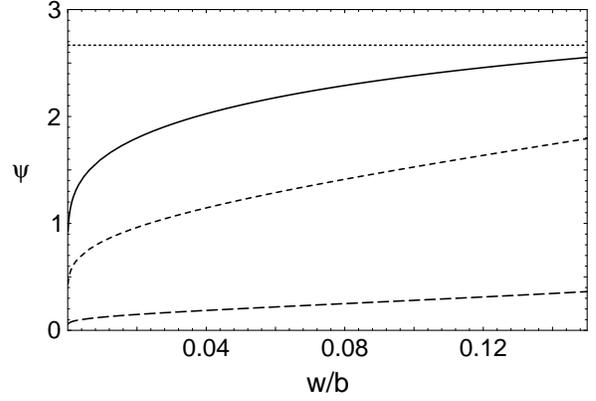

Fig. 1. Magnetic polarizability (in units of $b^3$) of a narrow annular cut versus its relative width $w/b$: solid line for (19), long-dashed line for octagon model (20), and short-dashed line for slot model (21). The dotted line shows the polarizability of the circular hole $\psi/b^3 = 8/3$.

which is still smaller than Eq. (19), see Fig. 1.

As seen from Fig. 1, the polarizability (19) becomes close to that of a circular hole for relatively narrow gaps, $w/b \geq 0.1$. The physical reason for this surprising result is that a tangential magnetic field very easily and deeply penetrate even through a very narrow annular gap in the thin wall. This distortion of the incident field creates a large effective magnetic dipole which is comparable to that due to the open hole with the same radius. As we shall see in Section IV, this is not the case in the electric problem. To find the applicability range for the analytical result of Eq. (19) and include thickness effects, we proceed below with a variational study of Eq. (5).

#### B. Wide Cut: Variational Approach

An elegant variational technique for polarizabilities has been developed in [5]. Multiplying Eq. (5) or (9) by $rg(r)$ and integrating over $r$, we convert it to the following variational form for the magnetic polarizability $\psi$

$$\frac{\pi b^3}{\psi} = \frac{\int_\rho^1 x dx \int_\rho^1 y dy \, g(x) K(x, y) g(y)}{\left[\int_\rho^1 x^2 \, dx \, g(x)\right]^2} \,, \qquad (22)$$

where kernel $K(x, y) = K_m(x, y)$ for $t = 0$ or $K(x, y) = K_m(x, y) + K_{mt}^{s,as}(x, y)$ for thick wall. A solution $g(x)$ of Eqs. (5) or (9) minimizes the RHS of Eq. (22). We are looking for a solution in the form of a series

$$g(x) = \sum_{n=0}^{\infty} c_n g_n(x) \qquad (23)$$

with unknown coefficients $c_n$. The choice of functions $g_n(x)$ is defined by the near-edge behavior of the solution. For the zero-thickness case we choose

$$g_0(x) = [(1 - x)(x - \rho)]^{-1/2} \,, \qquad (24)$$

$$g_k(x) = T_{k-1}\left(\frac{2x - \rho - 1}{1 - \rho}\right) \text{ for } k \geq 1 \,,$$

where $T_n(x)$ are Chebyshev's polynomials of the first kind. For a thick wall

$$g_0(x) = [(1-x)(x-\rho)]^{-1/3} , \qquad (25)$$
$$g_k(x) = C_{k-1}^{1/6}\left(\frac{2x-\rho-1}{1-\rho}\right) \text{ for } k \geq 1 ,$$

where $C_n^{1/6}(x)$ are Gegenbauer's polynomials. This choice of the polynomials is related to their orthogonality to the singular part $g_0(x)$ of the solution.

Denoting $d_n = \int_\rho^1 dx\, x^2 g_n(x)$ and $a_n = c_n d_n$, we define the matrix

$$K_{kn} = \int_\rho^1 x\,dx \int_\rho^1 y\,dy\, g_k(x)K(x,y)g_n(y)/(d_k d_n) , \quad (26)$$

and convert Eq. (22) into the following form

$$\frac{\pi b^3}{\psi} = \frac{\sum_{k,n} a_k K_{kn} a_n}{\left(\sum_n a_n\right)^2} . \qquad (27)$$

Following [5], one can prove that minimizing the RHS of Eq. (27) yields

$$\psi = \pi b^3 \sum_{k,n} \left(K^{-1}\right)_{kn} , \qquad (28)$$

where matrix $K^{-1}$ is the inverse of the matrix $K$ defined by Eq. (26). The further procedure is straightforward: $n$th iteration ($n = 0, 1, 2, \ldots$) corresponds to the matrix (26) truncated to the size $(n+1) \times (n+1)$. In the zeroth iteration the truncated matrix is merely a number $K_{00}$. All integrations and matrix inversions have been carried out using *Mathematica* [9].

Calculations show that for zero wall thickness only even terms of the series (23) contribute, i.e. $c_1 = c_3 = \ldots = 0$, and, effectively, one can use $g = g_0 + c_2 T_1 + c_4 T_3 + \ldots$, and squeeze matrix $K$ removing odd lines and rows. The results for $\psi$ ($\psi = \psi_{in} = \psi_{out}$ for $t = 0$) versus the cut width are shown in Fig. 2 (dashed line). One can see that the zeroth iteration, as well as the analytical solution (19), works well for narrow gaps, $w/b \leq 0.15$. The process practically converges in three iterations (effective 0,1,2) for the whole range of the cut width $0 \leq w/b \leq 1$.

For the case of a finite wall thickness, it is instructive to rewrite the variational equation (22) as

$$\frac{\psi}{\pi b^3} = \frac{\left[\int_\rho^1 x^2\, dx\, g(x)\right]^2}{D[g]} , \qquad (29)$$

where the functional $D[g]$ in the denominator is

$$D[g] \equiv \int_\rho^1 x\,dx \int_\rho^1 y\,dy\, g(x) K_m(x,y) g(y) + \qquad (30)$$
$$\sum_{n=1}^\infty \left[\int_\rho^1 x\,dx\, g(x) F_n(x)\right]^2 \frac{\tanh(\lambda_n t/2)}{\lambda_n} .$$

For the antisymmetric problem, tanh in Eq. (30) is replaced by coth. From Eqs. (29)-(30) one can easily see that due to the presence of the positive second term in $D[g]$ the magnetic polarizability for any $t > 0$ is reduced compared to that for zero-thickness case.

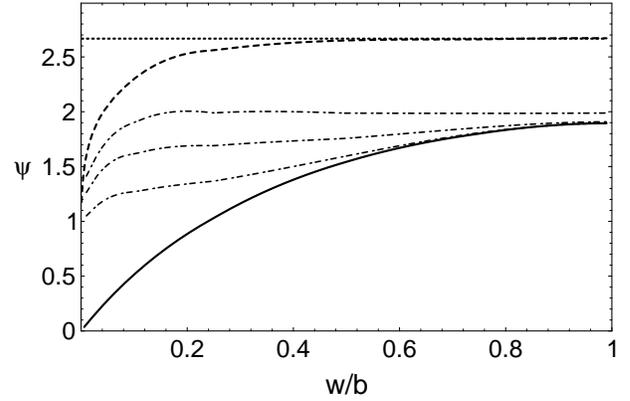

Fig. 2. Inside magnetic polarizability (in units of $b^3$) of an annular cut versus its relative width $w/b$ for thin (dashed) and thick (solid) wall. Three dash-dotted curves are for fixed ratio $t/w = 0.5; 1; 2$ (from top to bottom). The dotted line corresponds to the circular hole in a thin wall.

An asymptotic of $\psi$ for a narrow gap, $\delta = w/b \ll 1$, in a thick wall can be obtained easily using properties of eigenvalues: $\lambda_n b \to \pi(n-1)/\delta$ for $n \geq 2$, and $\lambda_1 b \simeq 1 + \delta/2$ when $\delta \to 0$. (In the opposite extreme, $\delta \to 1$, roots $\lambda_n b$ tend to the roots of $J_1'(x)$ from below.) When the wall is thick enough, i.e. $t > 2b$, one can neglect all terms except $n = 1$ in the sum in Eq. (30). From normalization condition for $F_n$ follows $F_1(x) \simeq \delta^{-1/2}$. Keeping only singular term $g_0(x)$ in series (23), and neglecting the term with $K_m$ in $D[g]$ since it is small compared to $[\int g(x) F_1(x)]^2 \propto \delta^{-1/3}$, we get after some algebra

$$\psi_s \simeq \psi_{as} \simeq \pi b^3 \delta = \pi b^2 w . \qquad (31)$$

It gives asymptotic $\psi_{in} = 2\pi b^2 w$ for a narrow annular gap in the thick wall. Comparison to the results of direct variational calculations for the thick wall in Fig. 2 (solid line) shows that this asymptotic works only for very small $w/b$, giving the initial slope of the curve in Fig. 2.

The variational calculations for the thick wall are similar to those for the zero-thickness case, except that one has to truncate the series in $n$ for the thickness-dependent part in Eq. (30). We have kept up to 6 terms in this series, and convergence was fast enough, requiring only up to 3 to 4 iterations. Again, for narrow gaps the process practically converges after the first iteration. Figure 3 shows the inside and outside magnetic polarizabilities versus the wall thickness for different values of the gap width. One can see that "thick-wall" asymptotics are reached approximately at $t/b = 2$. The outside polarizabilities decrease exponentially with thickness increase.

Figure 2 shows the inside magnetic polarizability as a function of the gap width for different wall thicknesses. One should mention that in the limit $w/b \to 1$ our results coincide with those obtained for a circular hole [5], e.g., $\psi_{in}(t \to \infty) = 0.71 \psi_{in}(t = 0)$.

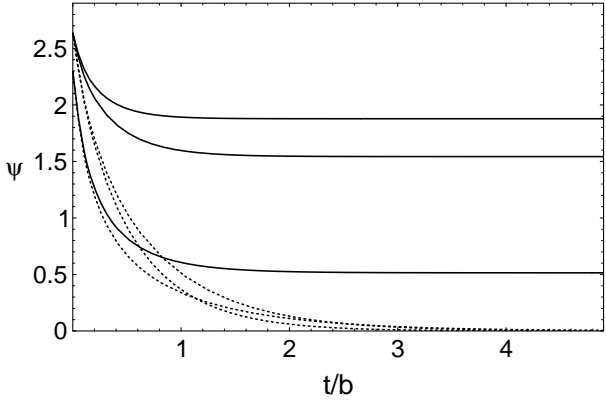

Fig. 3. Inside (solid) and outside (dotted) magnetic polarizability (in units of $b^3$) of an annular cut versus wall thickness for different relative widths $w/b = 0.1; 0.5; 0.9$ (from bottom to top).

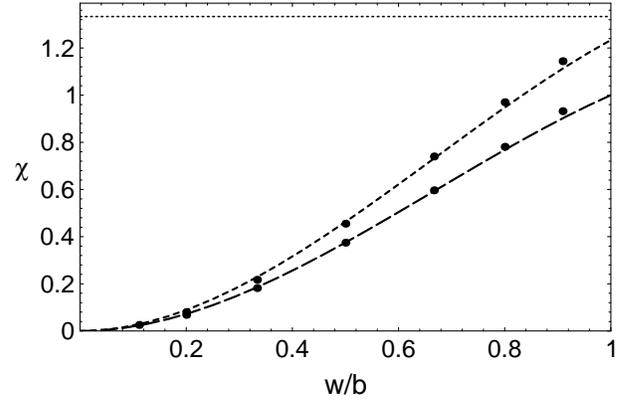

Fig. 4. Inside electric polarizability (in units of $b^3$) of an annular cut versus its relative width $w/b$: analytical estimates (32) for a thin wall (short-dashed) and (33) for a thick wall (long-dashed) and corresponding numerical results (thick dots). The dotted line is for the circular hole in a thin wall, $\chi/b^3 = 4/3$.

## IV. ELECTRIC PROBLEM

### A. Narrow Cut: Analytical Estimates

For a narrow annular cut $w \ll b$, the electric polarizability can be approximated by that of a narrow (yet bented) slot of width $w$ and length $\pi(b+a) \gg w$. The approximation is relevant as long as the width is small compared to the radius of curvature, and it gives $\chi \simeq \tilde{\chi}\pi(b+a)$, where $\tilde{\chi}$ denotes the electric polarizability per unit length of the slot. The value of $\tilde{\chi}$ can be obtained using conformal mapping for a 2-D electrostatic problem, and for two extreme cases the results are quite simple: $\tilde{\chi} = \pi w^2/8$ for zero wall thickness, and $\tilde{\chi} = w^2/\pi$ for a thick wall, $t \gg w$, see [7] and references therein. In this way, we have two simple analytical estimates for the electric polarizability of a narrow annular cut:
for a thin wall
$$\chi \simeq \pi^2 w^2 (b+a)/8 \,, \tag{32}$$
and for a thick wall
$$\chi_{in} \simeq w^2 (b+a) \,. \tag{33}$$

Obviously, for narrow gaps the electric polarizability is small compared to the magnetic one. The reason, from physical point of view, is that the normal electric field does not penetrate far enough through the narrow gap, unlike the tangential magnetic field on the parts of the annular cut which are parallel to its direction.

The outside electric polarizability of the gap in a thick wall is exponentially small. Taking a characteristic depth $w/\pi$ of the electric field penetration inside the gap and using Eq. (33) leads to the estimate
$$\chi_{out} \simeq w^2(b+a)\exp(-\pi t/w) \,. \tag{34}$$

### B. Wide Cut: Numerical Approach

Both the electro- and magnetostatic problems under consideration can be solved numerically. With boundary conditions which ensure a given homogeneous field far from the aperture plane, a static electric or magnetic potential could be computed using standard codes. Unfortunately, for the magnetic problem, as well as for an arbitrary-shaped aperture, this approach requires 3-D codes and cumbersome computations as a result. However, for the electric polarizability of an axisymmetric aperture the problem is effectively a 2-D one due to its axial symmetry. On the other hand, an application of the variational technique to the electrostatic problem under consideration is complicated since its zero-thickness kernel Eq. (15) is singular: direct numerical computations of variational integrals would be involved unless the integration is performed analytically (which is also very difficult in this case). That is why we choose the numerical approach applying the POISSON code [10].

For a numerical solution, we consider a conducting circular cylinder with the axis at $u = v = 0$, the radius $5b$, where $b$ is the aperture outer radius, and its base on the aperture plane. The cylinder "lid", which is at the distance about $10b$ from the aperture plane, is an equipotental surface, with its potential chosen to provide unit electric field near the surface. The potential of the aperture plane is fixed to be zero, and boundary conditions $d\phi/dn = 0$ on the side wall are imposed to force electric-field lines to be parallel to it. Imposing Neumann's boundary condition $d\phi/dn = 0$ inside the aperture, at $z = 0$, give us the symmetric problem for the potential $\phi$. Likewise, Dirichlet's boundary condition $\phi = 0$ in the aperture leads to the antisymmetric problem. Exploiting the axial symmetry of the problem, we use 2-D electrostatic code POISSON to solve for the potential $\phi(r, z)$. Then integrating $r\phi(r, z = t/2)$ from $a$ to $b$ gives us the electric polarizability, according to Eq. (16).

The results are shown in Fig. 4. Comparison of numerical results with analytical estimates (32) and (33) shows that these estimates work amazingly well even for very wide gaps. We intentionally did not interpolate the numerical results (dots) in Fig. 4, otherwise it would be very difficult to distinguish the numerical curves from those given by formulas (32)-(33); they practically overlap except in the region $w/b \geq 0.85$. Numerical results for finite wall thickness $t/w = 1$ and even $t/w = 0.5$ are very close to those for a very thick wall (the lower curve in Fig. 4).

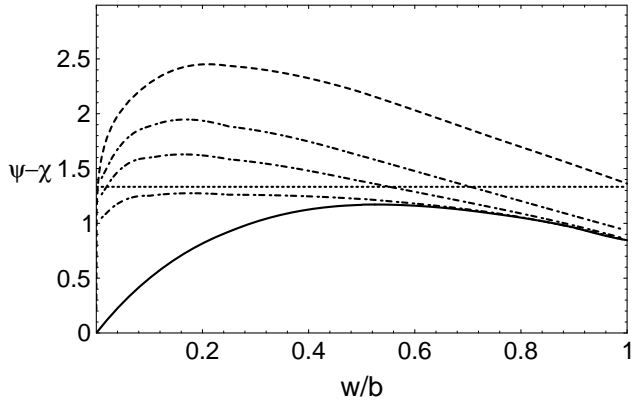

Fig. 5. Difference of inside polarizabilities (in units of $b^3$) of an annular cut versus its relative width $w/b$ for different thicknesses of the wall $t = 0$; $w/2$; $w$; $2w$, and $t \gg w$ (from top to bottom). The dotted line corresponds to the circular hole in a thin wall, $(\psi - \chi)/b^3 = 4/3$.

Estimate (34) of the outside electric polarizability coincides with numerical results within 10% for $w/b \leq 0.5$, and much better for narrow gaps. The thickness dependences of $\chi_{in}$ and $\chi_{out}$ for a fixed width is similar to those shown in Fig. 3 for the magnetic case. However, the thickness effect on the inside electric polarizability is weaker, as one can see from Fig. 4, and its asymptotic value is reached for thinner walls, at $t \geq w$.

## V. Beam Coupling Impedances

The beam-chamber coupling impedances can be obtained using formulas from [4] and polarizabilities found in Sect. III and IV. An annular cut of radius $b$ and width $w$ on the wall of a circular pipe of radius $r \gg b$ produces the longitudinal impedance

$$Z(\omega) = -\frac{iZ_0\omega(\psi_{in} - \chi_{in})}{8\pi^2 cr^2}, \quad (35)$$

where the difference $(\psi_{in} - \chi_{in})/b^3$ is plotted in Fig. 5 for different widths and wall thicknesses. As for other cross sections of the vacuum chamber, the real part of the impedance, and the transverse impedance, see [11] and references therein.

For the case of a narrow annular gap, $w \ll b$, on the thin wall, the magnetic polarizability dominates, according to (19) and (32), and Eq. (35) takes the following form

$$Z(\omega) \simeq -\frac{iZ_0\omega b^3}{8cr^2\left[\ln(32b/w) - 2\right]}. \quad (36)$$

Note that the impedance (36) of a narrow cut with $w/b > 0.05$ in a thin wall is larger than (but less than twice) that of a circular hole with radius $b$, and tends to the last one when $w \to b$.

The analytical expression (36) can be used as an upper estimate for the impedance of a button-type BPM. However, taking into account the wall thickness reduces the estimate significantly, cf. Fig. 5, so a much more accurate result can be obtained by making use of Eq. (35) and polarizabilities from Fig. 6. As an example, we estimate the broad-band impedance for BPMs of the PEP-II B-factory at SLAC and compare it with 3-D numerical simulations [12]. The BPM has 4 buttons of inner radius $a = 7.5$ mm, gap width $w = 1$ mm, at the distance $r = 30$ mm from the chamber axis. In fact, the PEP-II chamber has an octagonal cross section, but we approximate it by a circular pipe with radius 30 mm. While the wall thickness is not specified in [12], it is usually a few times larger than the gap width. The calculation according to (36) would give the inductance $L = 0.12$ nH per BPM ($Z = -i\omega L$) in a thin-wall approximation. The account of the wall thickness reduces this upper estimate, cf. Fig. 5: if the thickness is taken $t = 2w = 2$ mm, the result is $L = 0.06$ nH per BPM, and $L = 0.032$ nH for a very thick wall, $t \gg w$. The numerical result [12] is $L = 0.04$ nH per BPM, in a good agreement with our estimate for the case of a finite wall thickness.

## VI. Conclusions

The polarizabilities of an annular cut in a wall of any thickness are studied. The magnetic polarizability is calculated using the analytical and variational methods. To calculate the electric polarizability we applied the direct numerical approach. Combining different methods allows us to find the polarizabilities for different widths of the cut and to take into account the effects due to the wall thickness.

The results can be used for many applications of the aperture theory. As an example, the estimate for the coupling impedance of button-type BPMs is obtained.

## Acknowledgements

The author would like to thank Dr. R.L. Gluckstern and Dr. R.K. Cooper for useful discussions related to this work.

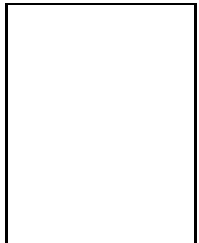

**Sergey Kurennoy** was born in Smolensk, Russia, in 1956. He graduated from the Moscow State University in 1980 with MS in Physics. From 1980 to 1983 he had a post-graduate course at the Theory Division of the Institute for High Energy Physics (IHEP), Serpukhov, Russia, working on nonperturbative problems in gauge field theories, and received his PhD in Theoretical and Mathematical Physics from the IHEP in 1985. He joined the UNK Department of the IHEP in 1983, and since then has worked in accelerator theory, mostly on beam dynamics and electromagnetic problems in accelerators. In 1992 Dr. Kurennoy joined the SSC Laboratory in Dallas, TX, and after the termination of the SSC project in 1994 he has been with the Physics Department, University of Maryland at College Park. His research interests include accelerator physics, nonlinear dynamics, and particle theory.